# Superconductivity, magnetism and crystal chemistry of $Ba_{1-x}K_xFe_2As_2$

Dirk Johrendt[a,*] and Rainer Pöttgen[b]

a   Department Chemie und Biochemie, Ludwig-Maximilians-Universität München,
    Butenandtstrasse 5–13 (Haus D), 81377 München, Germany
    E-mail: johrendt@lmu.de
b   Institut für Anorganische und Analytische Chemie and NRW Graduate School of
    Chemistry, Universität Münster, Corrensstrasse 30, 48149 Münster, Germany
    E-mail: pottgen@uni-muenster.de

## A B S T R A C T

$BaFe_2As_2$ is the parent compound of the '122' iron arsenide superconductors and crystallizes with the tetragonal $ThCr_2Si_2$ type structure, space group $I4/mmm$. A spin density wave transition at 140 K is accompanied by a symmetry reduction to space group $Fmmm$ and simultaneously by antiferromagnetic ordering. Hole-doping induces superconductivity in $Ba_{1-x}K_xFe_2As_2$ with a maximum $T_c$ of 38 K at $x \approx 0.4$. The upper critical fields approach 75 T with rather small anisotropy of $H_{c2}$. At low potassium concentrations ($x \leq 0.2$), superconductivity apparently co-exists with the orthorhombic distorted and magnetically ordered phase. At doping levels $x \geq 0.3$, the structural distortion and antiferromagnetic ordering is completely suppressed and the $T_c$ is maximized. No magnetically ordered domains could be detected in optimally doped $Ba_{1-x}K_xFe_2As_2$ ($x \geq 0.3$) by $^{57}Fe$-Mössbauer spectroscopy in contrast $\mu$SR results obtained with single crystals. The magnetic hyperfine interactions investigated by $^{57}Fe$ Mössbauer spectroscopy are discussed and compared to the $ZrCuSiAs$-type materials.

*Keywords:* $BaFe_2As_2$, Superconductivity, SDW anomaly, $^{57}Fe$-Mössbauer spectroscopy

*PACS:* 74.10.+v, 74.25.Ha, 74.25.Dw, 74.62.Bf, 33.45.+x

*Corresponding author. E-mail address*: johrendt@lmu.de (D. Johrendt)

## 1. Introduction

The quest for new superconductors is still a highly intuitive venture. Two decades after the discovery of the famous cuprates [1], the chemical ingredients to prepare new high-$T_c$ materials remain to be seen. While nature toughly defies disclosing the secret of the high-$T_c$ mechanism [2], no universally valid recipe regarding the elements involved or their spatial arrangement is in sight. Hence, it appears not surprising that most (not all) superconductors have been discovered by chance rather then by concept.

In February 2008, *Hideo Hosono* reported on superconductivity at 26 K in the iron arsenide $LaFeAs(O_{1-x}F_x)$ [3], while he was engaged in optoelectronic devices based on ZrCuSiAs-type compounds [4,5]. This discovery heralded a new era of super-conductivity based just on iron, unexpectedly. In fact, the very beginning was in 2006, when superconductivity in LaFePO [6] was reported, followed by LaNiPO [7,8]. However, the phosphides have been hardly noticed due to their low $T_c$'s of 3-7 K and the true breakthrough came with LaFeAsO. Within weeks, Chinese groups raised the $T_c$ up to 55 K in $SmFeAs(O_{1-x}F_x)$ [9] and it becomes apparent that iron arsenides represent a new class of high-$T_c$ superconductors, 22 years after the cuprates [10].

We had been involved in materials with ZrCuSiAs-type structure for a while [11,12] and in the crystal chemistry of the large family of compounds with the related $ThCr_2Si_2$-type structure for a longer time [13-16]. In this context, the potential of superconductivity in $ThCr_2Si_2$-type materials due to the electronic structure of the transition metal layers has been pointed out [17]. Since the ternary iron arsenides $AFe_2As_2$ ($A$ = Sr, Ba) and also $KFe_2As_2$ had been known to exist [18-20], the analogy to the superconductor LaFeAsO was immediately clear to us. We have demonstrated the very similar properties of the parent compounds LaFeAsO and $BaFe_2As_2$ [21] and we also succeeded in inducing superconductivity up to $T_c$ = 38 K by hole-doping in $Ba_{1-x}K_xFe_2As_2$ [22,23]. Reports on analogous compounds $Sr_{1-x}K_xFe_2As_2$ [24], $Ca_{1-x}Na_xFe_2As_2$ [25] and $Eu_{1-x}K_xFe_2As_2$ [26] followed quickly. Shortly after this, superconductivity under pressure has been found in the undoped parent compounds $BaFe_2As_2$ and $SrFe_2As_2$ [27] and possibly in $EuFe_2As_2$ [28]. First reports about pressure induced superconductivity in $CaFe_2As_2$ were recently contradicted [29]. This is not surprising, since $CaFe_2As_2$ can be compressed to a three dimensional structure with As−As bonds (improperly called 'collapsed phase') [30], as it is well known from other ThCr$_2$Si$_2$-type compounds [13], but unique in the class of 122-superconductors. Thus, CaFe$_2$As$_2$ is not an appropriate model system of 122-family [31].

The critical temperatures of 122-superconductore have reached 38 K and are thus lower than 55 K in the $R$FeAsO (1111) family. On the other side, phase pure samples are easier to obtain and do not suffer from synthetic problems associated with the control of oxygen or fluorine, and also single crystals are readily accessible [32]. Since also the underlying physics are likely the same, the 122-superconductors have become subject to very extensive studies and much of the progress in this fascinating new field is based on these compounds. In this article, we will review the crystal chemistry, structural features, magnetism and superconductivity of BaFe$_2$As$_2$ and the solid solution Ba$_{1-x}$K$_x$Fe$_2$As$_2$.

## 2. Synthesis

Powder samples of the ternary parent compound BaFe$_2$As$_2$ can be synthesized by heating stoichiometric mixtures of the elements at 650-900°C in alumina crucibles under atmospheres of purified argon. Barium metal should be purified by distillation and arsenic by sublimation before use to avoid oxygen contamination. Repeated homogenization and annealing is necessary to obtain pure samples, where foreign phases contribute less than one percent. This is the typical amount scarcely detectable by x-ray powder diffraction, but one should keep in mind that even smaller contaminations can affect physical property measurements. In particular, small traces of ferromagnetic phases (like Fe metal) strongly interfere with the weak magnetism of BaFe$_2$As$_2$. Also iron arsenides like antiferromagnetic FeAs ($T_N \approx 77$ K) and Fe$_2$As ($T_N \approx 50$ K) can cause misinterpretations.

Potassium-doped powder samples Ba$_{1-x}$K$_x$Fe$_2$As$_2$ can also be synthesized from the elements, but the alkaline metal cause problems due to vaporization and reaction with the alumina crucible as well as the silica tubes. These can be avoided by minimizing the free gas volume in the tubes or by allowing the potassium to pre-react with arsenic at lower temperatures. The latter allows the use of closed niobium or tantalum ampoules, which prevents any loss of potassium by vaporization. The pre-reacted products were homogenized, cold pressed to pellets and subsequently annealed until homogeneity is achieved.

The synthesis of single crystals is by far less straightforward then initially supposed. The use of a tin flux leads to incorporation of tin in $BaFe_2As_2$ according to $Ba_{1-x}Sn_xFe_2As_2$ with $x \approx 0.05$, which slightly affects the properties [33]. The FeAs self-flux method [34] avoids this problem and appears to be currently the best method to obtain crystals of undoped $BaFe_2As_2$. However, one has to pay attention to a potential contamination of the crystals by inclusions of binary iron arsenides. The use of the self flux method to grow single crystals of $Ba_{1-x}K_xFe_2As_2$ is very problematic. The potassium incorporation is hardly controlled and mostly an excess of K-metal is used. Thus, such widely used 'high-quality' crystals [35] can never be homogeneous as a consequence of the phase diagram, because the potassium concentration in the flux is varying during the crystal growth. K-doped crystals out of tin fluxes [36] will suffer from the same problem in addition to the uncontrolled tin incorporation. Thus, the growth of high-quality homogeneous single crystals of $Ba_{1-x}K_xFe_2As_2$ is an unresolved problem and requires further experimental efforts in the near future.

## 3. Crystal Structure, Chemical Bonding and SDW-Transition

$BaFe_2As_2$ crystallizes in the tetragonal $ThCr_2Si_2$-type structure [37] with two formula units per unit cell (space group $I4/mmm$, $a = 396.25(1)$ pm, $c = 1301.68(3)$ pm, $Z = 2$) [21]. The crystal structure is depicted in Figure 1 and easily described as layers formed by edge-sharing $FeAs_{4/4}$ tetrahedra perpendicular to [001], separated by barium ions. A mirror plane is located between each layer in contrast to LaFeAsO, where each layer has the same orientation. The polar but covalent Fe−As bonds ($d_{Fe−As} = 240.3$ pm) within the layers are the by far strongest ones, followed by the rather ionic Ba−As contacts ($d_{Ba−As} = 338.2$ pm). Each barium atom has eight arsenic neighbors, which form a tetragonal prism. The distance of the arsenic atoms between the layers ($d_{As−As} = 378.8$ pm) is much too long for any kind of interaction.

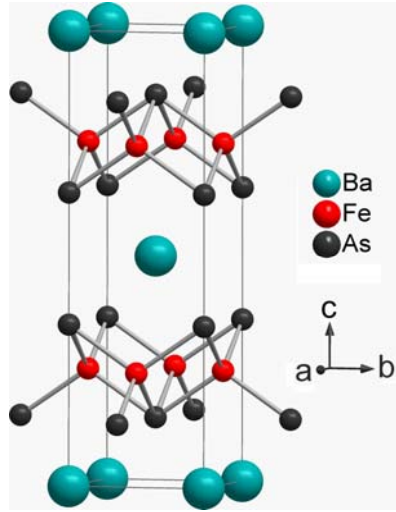

**Figure 1**. Crystal structure of $BaFe_2As_2$

As we pointed out earlier [17], metal-metal bonding within the layers plays an important role in the properties of $ThCr_2Si_2$-type compounds. This is especially the case in such compounds, where the Fermi level crosses the bands that originate from the direct $dd\sigma$- or $dd\pi$-type overlap. Assuming the iron atoms are in the $Fe^{2+}$ state ($3d^6$), the $d$-shell is more than half-filled and Fe–Fe anti-bonding states should be at least partially occupied. The lowest lying are the $dd\pi^*$-bands, made by the overlap of the Fe-$3d_{x^2-y^2}$ orbitals. Hence, the undoubtedly present Fe–Fe bonds ($d_{Fe-Fe} = 280.2$ pm) in $BaFe_2As_2$ are slightly weakened, but more important, these bands are less dispersed and mainly responsible for the magnetic properties. $BaFe_2As_2$ is on the verge of a magnetic instability. Any change of the $dd\pi^*$-overlap (by larger or smaller Fe-Fe distances) will either broaden or flatten the $d_{x^2-y^2}$ band, leading consequently to a nonmagnetic or magnetic state, respectively. By taking the chemical bonding into account, we have the typical Peierls-scenario, where the total energy can be lowered by a structural distortion, *i.e.* a reduction of symmetry. Thus, the magnetic and structural degrees of freedom are intimately coupled.

In fact, $BaFe_2As_2$ exhibits a structural and magnetic phase transition at $T_o = 140$ K, which is associated with anomalies in the specific heat, electrical resistivity and magnetic susceptibility [21]. Below $T_o$, the tetragonal symmetry is reduced to orthorhombic (space group *Fmmm*, *Z* = 4, *a* = 561.46(1) pm, *b* = 557.42(1) pm, *c* = 1294.53(1) pm), which leads to a small splitting of the Fe–Fe bond lengths to 280.7(2) and 278.7(1) pm, respectively. The temperature dependence of the lattice parameters is shown in Figure 2.

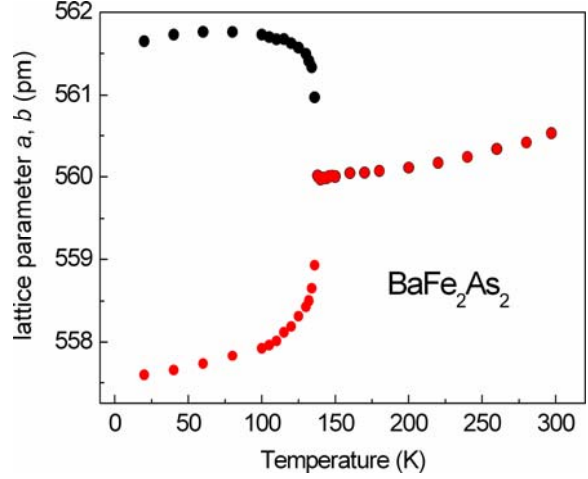

**Figure 2.** Lattice parameters of $BaFe_2As_2$

Antiferromagnetic ordering appears simultaneously with the structural distortion ($T_N \approx T_o$). Strong magnetic correlations also exist above the transition and with respect to the metallic nature of the compound, this is attributed to a spin-density-wave (SDW) state as known from other antiferromagnetic metals like chromium and its alloys [38]. Even though in iron arsenides the onset of magnetic ordering is coupled to a structural distortion, the term SDW instability is commonly used, which conceals the structural nature of the transition. This kind of SDW instability has been first observed in LaFeAsO [39], where the magnetic order emerges below the structural distortion temperature ($T_N < T_o$), and later in the parent compounds of all iron arsenide superconductors without exception. From the structural point of view, the distortion mechanism may be phonon mode-softening when approaching $T_o$, and would then be a typical second order transition. Interestingly, exactly this kind of second order transition is well established in the parent compound of the cuprates $La_2CuO_4$ at $T_o$ = 520 K ($T_N \approx$ 300 K) [40]. The SDW anomaly is certainly one important common feature of the iron arsenide superconductors and has become subject to extensive studies [21,25,41-45]. However, such SDW or CDW (charge-density-wave) phenomena are by no means specific features of iron arsenides, but in fact known from other superconductors with competing instabilities like cuprates [46,47], $UGe_2$ [48], $NbSe_3$ [49] or organic materials like $(TMTSF)_2PF_6$ [50].

The crystal structure and SDW transition of $BaFe_2As_2$ is strongly affected by hole-doping with potassium. Within the solid solution $Ba_{1-x}K_xFe_2As_2$, we found continuous linear changes of the lattice parameters at room temperature [23]. As shown in Figure 3, the structural changes mainly affects the Fe–Fe bond lengths and the As-Fe-As angle ($\varepsilon$) within the layers. Since the unit cell volume remains constant, the driving force is rather an electronic than a geometric effect.

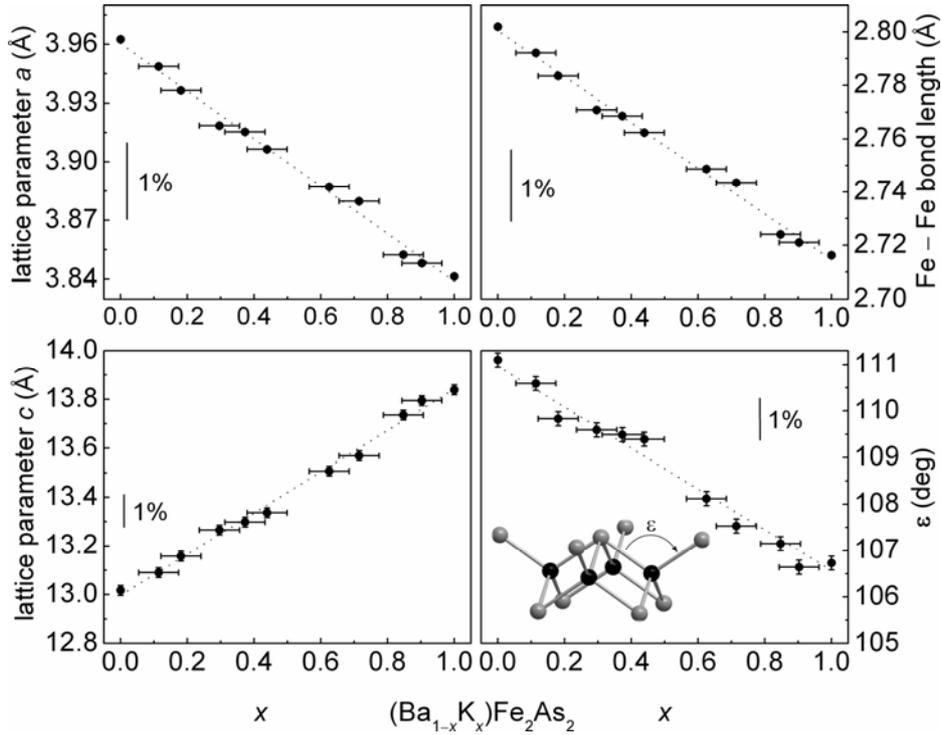

**Figure 3**. Lattice parameters, Fe–Fe distances and As-Fe-As angles in $Ba_{1-x}K_xFe_2As_2$ [23]

This can be understood within the metal-metal bonding scenario as explained above. If the electron count decreases by potassium doping, the $dd\pi^*$ antibonding bands close to the Fermi level become less occupied and thus the Fe−Fe bonds become stronger, *viz.* shorter. This effect contracts the lattice parameter *a*, and in order to keep the Fe−As and Ba−As distances constant (strong bonds), ε decreases and the unit cell elongates in *c*. It is interesting to note that ε is just close to the ideal tetrahedral angle 109.5° when approximately half of the barium atoms are substituted by potassium. This coincides with the maximum $T_c$ of 38 K around *x* = 0.4-0.5 and it has been speculated, whether this ideal tetrahedral symmetry may be connected with the highest $T_c$.

The structural distortion in $Ba_{1-x}K_xFe_2As_2$ remains present at low doping and was clearly detected up to *x* = 0.2 [23,51]. Figure 4 displays the temperature dependency of the lattice parameters at different doping levels. The transition temperature $T_o$ shifts to lower values, the transition becomes less sharp and the splitting of the lattice parameters decreases. No anomaly is evident at *x* = 0.3 and higher potassium concentrations.

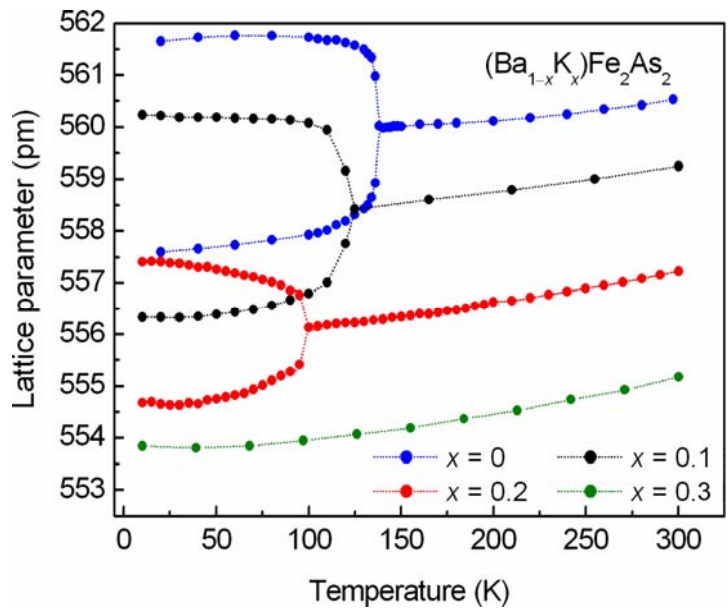

**Figure 4**. Lattice parameters of the series $Ba_{1-x}K_xFe_2As_2$ (*x* = 0 − 0.3) [51]

## 4. Superconductivity and Phase Diagram

Superconductivity in hole-doped 122-compounds was first reported in $Ba_{0.6}K_{0.4}Fe_2As_2$ with $T_c = 38$ K [22]. This is still the highest critical temperature among the 122-family up now and may be due to the largest interlayer distance, which seemingly scales with $T_c$ in iron arsenides. Figure 5 shows some details of the resistivity transition, which is not extremely sharp, because of a certain inhomogeneity in the potassium distribution. Later it was shown, that the superconducting dome has a relatively broad maximum around $x = 0.4$-$0.5$ [23,52], therefore small variations in the potassium distribution are not expected to cause significant changes of $T_c$. $Ba_{0.6}K_{0.4}Fe_2As_2$ is a bulk superconductor, as clearly proved by the magnetic susceptibility measurements as depicted in Figure 6. Note the Meißner-fraction of $\approx$ 60% at 0.5 mT.

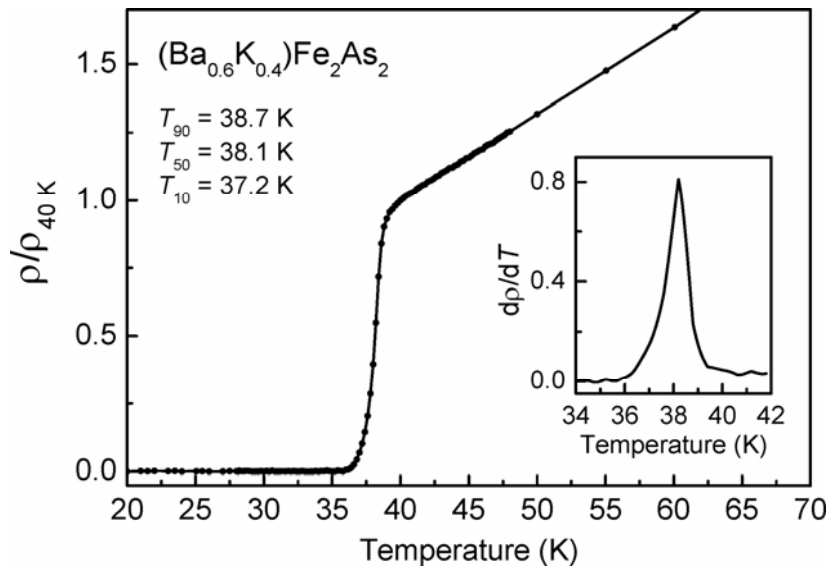

**Figure 5**. Resistivity transition of $Ba_{0.6}K_{0.4}Fe_2As_2$ [22]

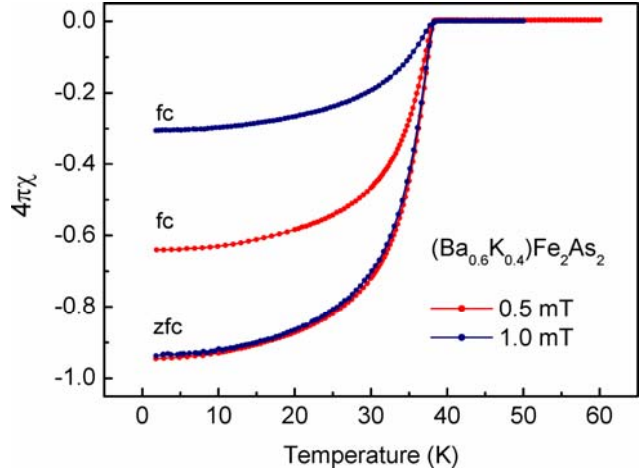

**Figure 6**. Magnetic susceptibility of $Ba_{0.6}K_{0.4}Fe_2As_2$ [22].

The upper critical fields and also the critical field anisotropy $\gamma = H_{c2}^{H\perp c}/H_{c2}^{H//c}$ seem to be not yet fully converged. Recently published data (extrapolated low field data are not reliable) of $Ba_{0.55}K_{0.45}Fe_2As_2$ show anisotropic $H_{c2}(T)$ up to 60 T and a nonlinear temperature dependence of the $H_{c2}$ anisotropy of $\gamma = 3.5$ close to $T_c$ that decreases to $\gamma = 1.2$ at 14 K [53]. Comparable small anisotropies at low temperature were reported in Ref. [54]. $H_{c2}(0)$ of this compound is expected to approach fields of 75 T or above.

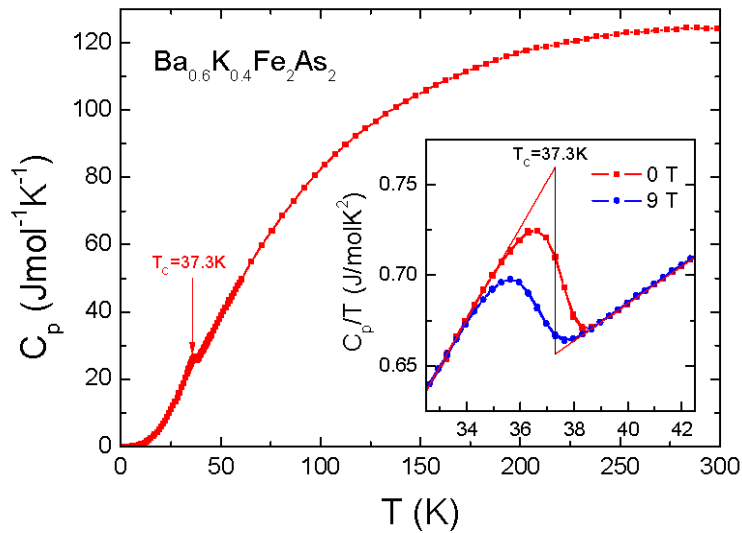

**Figure 7**. Specific heat of $Ba_{0.6}K_{0.4}Fe_2As_2$ [51].

The specific heat of $Ba_{0.6}K_{0.4}Fe_2As_2$ measured at zero field and 9 T is shown in Figure 7. The SDW anomaly is no longer present, but the superconducting is clearly discernible. The $T_C$ in zero field is 37.3 K by the entropy conserving construction and the change of the specific heat at the transition is estimated as $\Delta C/T_c = 0.1$ J/molK [51]. $T_c$ is shifted by only 1 K in at 9 T, reflecting the large upper critical field.

As the structural and magnetic transitions of $BaFe_2As_2$ begin to diminish by hole doping, superconductivity emerges in $Ba_{1-x}K_xFe_2As_2$ even at small potassium contents of $x = 0.1$ ($T_c \approx 3$ K), where the SDW anomaly is still present [23] (Figure 8). The resistivity of the $x = 0.2$ sample still shows the typical bump caused by the phase transition (smeared over a wider temperature range now), and a superconducting transition at 24 K. When the doping concentration reaches 0.3 (checked by EDX and chemical analysis), the SDW transition is completely suppressed and the $T_c$ increases to 33 K [51]. From our x-ray structure determinations [23], resistivity data and $^{57}$Fe-Mössbauer measurements [51], we compiled the phase diagram of $Ba_{1-x}K_xFe_2As_2$ as shown in Figure 9. The overlap of the orthorhombically distorted antiferromagnetic phase with the superconducting dome is apparent and reminiscent of single layer cuprate superconductors like $La_{2-x}Sr_xCuO_4$ [40].

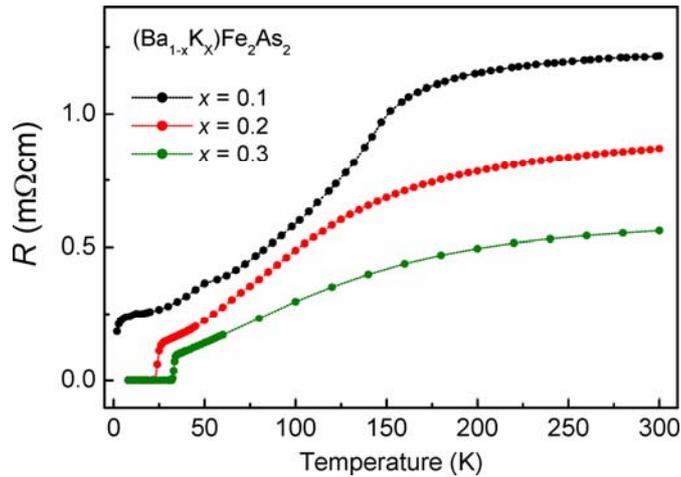

**Figure 8.** Resistance of $Ba_{1-x}K_xFe_2As_2$ ($x = 0.1$-0.3) from [51]

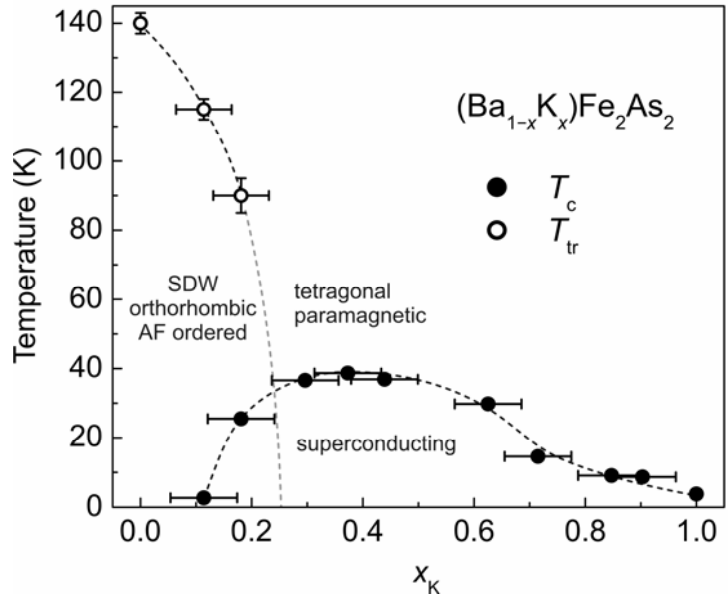

**Figure 9**. Phase diagram of $Ba_{1-x}K_xFe_2As_2$

Almost the same phase diagram has been constructed from neutron diffraction data [52]. However, the discussion about the co-existence of the magnetically ordered phase with superconductivity is still going on. Several $\mu$SR studies suggested mesoscopic phase separations of magnetically ordered and superconducting domains in almost optimally doped $Ba_{1-x}K_xFe_2As_2$ single crystals [55-57]. On the other hand, our [57]Fe-Mössbauer results, obtained with powder samples, showed magnetic ordering without paramagnetic components coexisting with superconductivity at $x = 0.2$ ($T_c = 24$ K) and completely non-magnetic phases at $x = 0.3$ ($T_c = 33$ K) and $x = 0.4$ ($T_c = 38$ K), respectively. Since our data are in agreement with the neutron powder diffraction study [52], we suggest that the single crystals used for the $\mu$SR experiments may have been inhomogeneous for the reasons explained in chapter 2. However, from our point of view, superconductivity and antiferromagnetic ordering coexists at least in the under-doped region of $Ba_{1-x}K_xFe_2As_2$ ($x \leq 0.2$).

## 5. Neutron Diffraction

The magnetic nature of the structural phase transition in $BaFe_2As_2$ has been first demonstrated by [57]Fe-Mössbauer spectra measured above and below $T_o$ [21]. The ordered magnetic moment was estimated to $0.4\mu_B$/Fe. A following neutron diffraction study with polycrystalline $BaFe_2As_2$ illuminated the complete spin structure [44], which is largely the same as in LaFeAsO and illustrated in Figure 10. The fundamental magnetic wave vector is $\mathbf{q} = (101)_O$, thus the magnetic moments are anti-parallel aligned to the longer orthorhombic *a*-axis (non-standard setting) and also antiferromagnetically along the *c*-axis.

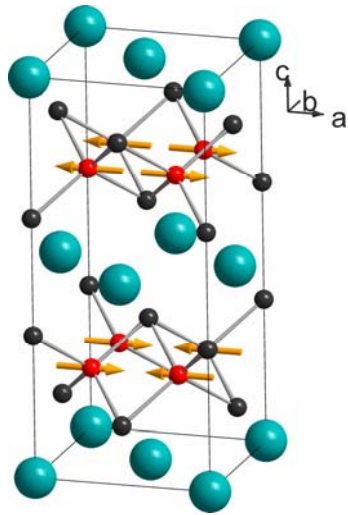

**Figure 10.** Magnetic structure of $BaFe_2As_2$. The magnetic moment is 0.87(3) $\mu_B$/Fe [44]

$BaFe_2As_2$ single crystals have also been studied and confirmed the powder data. The first single crystal experiments were conducted with crystals grown from a tin flux [33]. Even though these specimens showed substantially different phase transition and magnetic ordering temperatures due to tin incorporation, the determined magnetic structure agrees completely with the powder results in Ref. [44]. A later undertaken study with tin-free crystals [58] confirmed the spin structure again and proved, that the tin incorporation affected the transition temperatures, but not the magnetic structure.

Neutron diffraction experiments of the solid solution $Ba_{1-x}K_xFe_2As_2$ showed a co-existence of magnetic ordering with superconductivity at doping levels $x < 0.4$ [52]. In our experiments [51], the structural distortion and magnetic ordering is completely absent already at $x = 0.3$. This discrepancy may be due to different preparation techniques. Beyond these structural investigations, inelastic neutron scattering has been used to investigate phonon dynamics in $BaFe_2As_2$ [59] and to detect a resonant magnetic excitation in the superconducting phase of $Ba_{0.6}K_{0.4}Fe_2As_2$ [60]. This important result demonstrates that the superconducting energy gap has unconventional symmetry in the iron arsenide superconductors, as in the cuprates.

## 6. [57]Fe Mössbauer Spectroscopy

A valuable local probe for the investigation of oxidation states and magnetic hyperfine interactions in iron arsenides is [57]Fe Mössbauer spectroscopy. Besides the ZrCuSiAs type superconductors LaFePO [61] and $R$FeAsO ($RE$ = La, Ce, Pr, Nd) [62-67], 'LaFeAs' [68], $LaFeAsO_{1-x}F_x$ [69], and SrFeAsF [45] also the $ThCr_2Si_2$ type arsenides $A$Fe$_2$As$_2$ ($A$ = Ba, Sr, Eu, K) [21,41,51,68] have thoroughly been studied. Here we focus on the [57]Fe data of the 122-compounds $Ba_{1-x}K_xFe_2As_2$ [23].

In Figure 11 we present the 4.2 K [57]Fe-Mössbauer spectra of various samples $Ba_{1-x}K_xFe_2As_2$. The corresponding fitting parameters for these spectra are listed in Table 1. In agreement with the tetragonal $ThCr_2Si_2$-type crystal structure, the spectra show only one spectral component. As a consequence of the lower valence electron concentration, we observe a slight decrease of the isomer shift in going from $BaFe_2As_2$ to $KFe_2As_2$. The isomer shifts of the $Ba_{1-x}K_xFe_2As_2$ samples are close to the ones observed for the ZrCuSiAs type materials [61-69], $EuFe_2As_2$ [68] and the phosphides $R$Fe$_2$P$_2$ ($R$ = Ca, Sr, Ba, La, Ce, Pr, Eu) [70] with almost divalent iron.

Due to the non-cubic site symmetry of the iron atoms, all spectra show quadrupole splitting. $BaFe_2As_2$, $Ba_{0.9}K_{0.1}Fe_2As_2$, and $Ba_{0.8}K_{0.2}Fe_2As_2$ show full magnetic hyperfine field splitting (Figure 11) at 4.2 K with hyperfine fields slightly larger than 5 T (this corresponds approximately to a magnetic moment of 0.4 $\mu_B$ per iron atom.). These samples show no superconducting transition. Already in $Ba_{0.7}K_{0.3}Fe_2As_2$ we observe complete suppression of the SDW anomaly. Consequently, the [57]Fe Mössbauer spectra of $Ba_{0.7}K_{0.3}Fe_2As_2$ and $Ba_{0.6}K_{0.4}Fe_2As_2$ show single signals and no magnetic splitting. This in contrast to various studies on almost optimally doped single crystals of $Ba_{0.6}K_{0.4}Fe_2As_2$, which showed mesoscopic phase separations in antiferromagnetically ordered and non-magnetic superconducting domains [55-57]. Most likely these crystals had strongly inhomogeneous potassium contribution or contained small amounts of the impurity phase FeAs which orders antiferromagnetically at 77 K [71]. The other borderline phase, $KFe_2As_2$ [20,23], is a normal metal which shows a superconducting transition at $T_c = 3.8$ K. The 4.2 K [57]Fe Mössbauer spectrum (Figure 11) shows a single signal which is subject to weak quadrupole splitting. For further details and the complete [57]Fe Mössbauer spectroscopic study of the $Ba_{1-x}K_xFe_2As_2$ samples over the whole temperature range we refer to the original work [21,41,51].

**Table 1**

[57]Fe Mössbauer spectroscopic data of $BaFe_2As_2$, $Ba_{0.9}K_{0.1}Fe_2As_2$, $Ba_{0.8}K_{0.2}Fe_2As_2$, and $KFe_2As_2$ at 4.2 K. ($\delta$), isomer shift; ($\Gamma$), experimental line width, ($\Delta E_Q$), quadrupole splitting parameter, ($B_{hf}$), magnetic hyperfine field. The data refer to a [57]Co/Rh source were taken from references [21, 41, 51].

| Compound | $\delta$ (mm s$^{-1}$) | $\Delta E_Q$ (mm s$^{-1}$) | $\Gamma$ (mm s$^{-1}$) | $B_{Hf}$ (T) |
|---|---|---|---|---|
| $SrFe_2As_2$ | 0.47 | −0.09 | 0.37 | 8.91 |
| $BaFe_2As_2$ | 0.44 | −0.04 | 0.25 | 5.47 |
| $Ba_{0.9}K_{0.1}Fe_2As_2$ | 0.44 | −0.04 | 0.36 | 5.57 |
| $Ba_{0.8}K_{0.2}Fe_2As_2$ | 0.43 | −0.04 | 0.31 | 5.07 |
| $Ba_{0.7}K_{0.3}Fe_2As_2$ | 0.41 | −0.02 | 0.47 | |
| $Ba_{0.6}K_{0.4}Fe_2As_2$ | 0.39 | −0.10 | 0.35 | |
| $KFe_2As_2$ | 0.34 | −0.09 | 0.40 | |

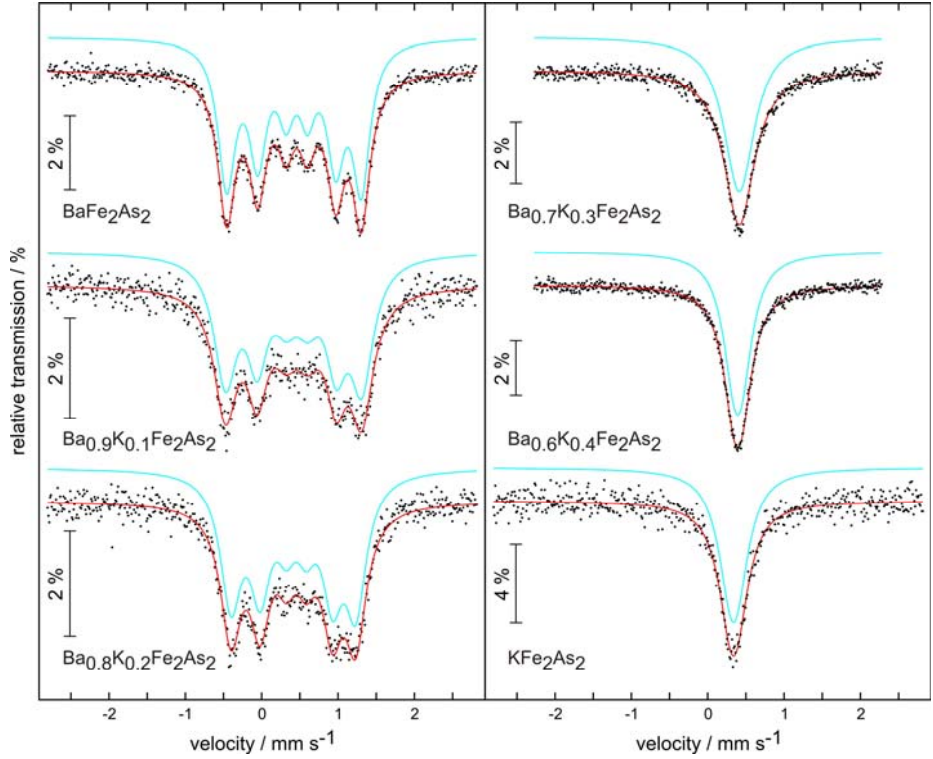

**Figure 11.** Experimental and simulated $^{57}$Fe Mössbauer spectra of samples from the solid solution $Ba_{1-x}K_xFe_2As_2$ measured at 4.2 K.

$^{57}$Fe-Mössbauer spectroscopy is also useful to detect foreign phases in iron arsenide samples. *Nowik* and *Felner* have systematically studied the influence of $Fe_2As$, $FeAs$, and $FeAs_2$ impurities on the SDW transitions and the superconducting properties [64] and it is also possible to distinguish stoichiometric LaFeAsO from oxygen deficient $LaFeAsO_{1-x}$ [63]. Decomposition of $SrFe_2As_2$ and $BaFe_2As_2$ leaves binary FeAs as one of the hydrolysis products. In Figure 12 we present a typical $^{57}$Fe Mössbauer spectrum (77 K data) from a decomposed $SrFe_2As_2$ sample which contains equal amounts of $SrFe_2As_2$ and FeAs. The fitting parameters for both phases ($\delta = 0.44(1)$ mm/s, $\Delta E_Q = -0.09(1)$ mm/s, $\Gamma = 0.30(1)$ mm/s and $B_{hf} = 8.70(2)$ T for the $SrFe_2As_2$ and $\delta = 0.47(1)$ mm/s, $\Delta E_Q = 0.65(1)$ mm/s, $\Gamma = 0.26(1)$ mm/s for the FeAs fraction) are in close agreement with the data originally reported for pure $SrFe_2As_2$ [41] and FeAs [72,73].

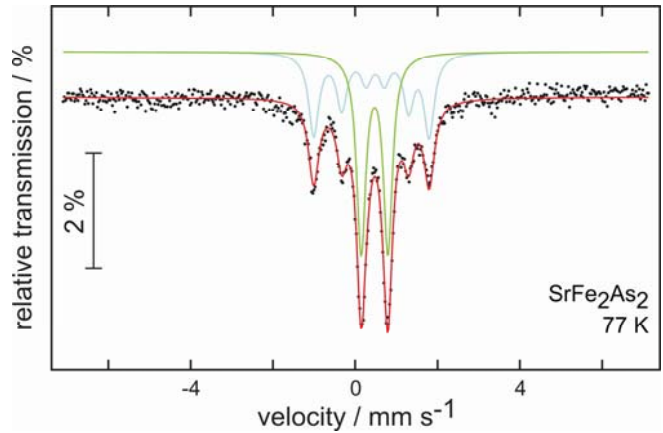

**Figure 12.** Experimental and simulated $^{57}$Fe Mössbauer spectrum of a partially decomposed SrFe$_2$As$_2$ sample at 77 K, containing equal amounts of SrFe$_2$As$_2$ and FeAs.

## 5. Summary

The discovery of high-$T_c$ superconductivity in Ba$_{1-x}$K$_x$Fe$_2$As$_2$ has stimulated extensive studies and a considerable piece of the progress in the field of iron arsenide superconductors is based this and the related 122-compounds. In comparison to the 1111-family with higher $T_c$'s, the 122-compounds are easier to obtain and also single crystals are accessible. However, in face of the huge number of experiments that have already been done, the sample quality and especially the homogeneity of the single crystals are still serious problems.

The parent compound BaFe$_2$As$_2$ crystallizes in the tetragonal ThCr$_2$Si$_2$-type structure and exhibits a structural and magnetic phase transition (SDW anomaly) at 140 K, connected with the onset of antiferromagnetic ordering. Doping of the barium atoms by potassium induces superconductivity in Ba$_{1-x}$K$_x$Fe$_2$As$_2$ with a maximum $T_c$ of 38 K at $x \approx 0.4$. The upper critical fields are expected to approach 75 T, whereas the anisotropy of $H_{c2}$ seems to be small ($\gamma \approx$ 1-3) and temperature dependent.

At low potassium concentrations ($x \leq 0.2$), superconductivity co-exists with the orthorhombic distorted and magnetically ordered phase according to neutron powder diffraction and $^{57}$Fe-Mössbauer spectroscopy. At doping levels $x \geq 0.3$, the structural distortion and antiferromagnetic ordering is completely suppressed and the $T_c$ is maximized. No magnetically ordered domains could be detected in optimally doped $Ba_{1-x}K_xFe_2As_2$ ($x \geq 0.3$) by $^{57}$Fe-Mössbauer spectroscopy in contrast to recent $\mu$SR results obtained with single crystals.

**Acknowledgements**

We are indebted to our coworkers Marianne Rotter, Marcus Tegel, Michael Pangerl, Veronika Weiß (LMU München), Falko M. Schappacher, Inga Schellenberg and Wilfried Hermes (WWU Münster) for their commitment, enthusiasm and many fruitful discussions. We thank Dr. Joachim Deisenhofer (Universität Augsburg) for specific heat and susceptibility measurements. This work was supported by the Deutsche Forschungsgemeinschaft.